\documentclass[british,12pt]{iopart}
\usepackage[T1]{fontenc}
\usepackage[latin9]{inputenc}
\usepackage{amstext}
\usepackage{amssymb}
\usepackage{graphicx}

\makeatletter
\usepackage{iopams}
\usepackage{setstack}

\usepackage{iopams}

\makeatother

\usepackage{babel}
\begin{document}
\title[Spectral description of the dynamics of interacting bosons in disordered lattices]{Spectral
description of the dynamics of ultracold interacting bosons in disordered
lattices}

\author{B Vermersch and J C Garreau}

\address{Universit\'e Lille 1 Sciences et Technologies, CNRS; F-59655
Villeneuve d'Ascq Cedex, France. \eads{http://www.phlam.univ-lille1.fr/atfr/cq}}

\begin{abstract}We study the dynamics of a nonlinear one-dimensional
disordered system from a spectral point of view. The spectral entropy
and the Lyapunov exponent are extracted from the short time dynamics,
and shown to give a pertinent characterization of the different dynamical
regimes. The chaotic and self-trapped regimes are governed by log-normal
laws whose origin is traced to the exponential shape of the eigenstates
of the linear problem. These quantities satisfy scaling laws depending
on the initial state and explain the system behaviour at longer times.\end{abstract} 


\section{Introduction}

The motion of non-interacting particles in disordered lattices has
been intensely studied over the last decades. In one and two dimensions,
and for a sufficient amount of disorder in three dimensions, it has
been shown that the spreading of quantum wavepackets is suppressed,
a phenomenon known as Anderson localization~\cite{Anderson:LocAnderson:PR58,Abrahams:Scaling:PRL79}.
However, the celebrated Anderson model that leads to this prediction
is a highly simplified model which in particular neglects particle-particle
interactions, so that a crucial question is: Do interactions destroy
(or, on the contrary, enhance) Anderson localization?

A burst of interest on the subject was recently driven by the demonstration
that these questions could be studied experimentally and theoretically
with an unprecedented degree of cleanness and precision using ultracold
atoms. This has produced an impressive number of results, concerning
both Anderson localization (AL)~\cite{Moore:AtomOpticsRealizationQKR:PRL95,Billy:AndersonBEC1D:N08,Roati:AubryAndreBEC1D:N08,Kondov:ThreeDimensionalAnderson:S11,Jendrzejewski:AndersonLoc3D:NP12,Lignier:Reversibility:PRL05}
and the Anderson transition~\cite{Chabe:Anderson:PRL08,Lemarie:AndersonLong:PRA09,Lemarie:CriticalStateAndersonTransition:PRL10,Lopez:ExperimentalTestOfUniversality:PRL12},
observed in 3D systems. Moreover, systems of ultracold bosons turned
out to be very well modeled by mean-field approaches~\cite{Stringari:BECRevTh:RMP99,Bloch:ManyBodyUltracold:RMP08},
in contrast to fermionic systems where such a simplification of the
corresponding many-body problem is not possible.

From the theoretical and numerical point of view, ultracold bosons
in a 1D optical disordered lattice can be described rather realistically
by a simple generalization of the original Anderson model including
a nonlinear term taking into account interactions in a mean-field
description. Numerical studies of the model suggest that the long
term motion of the particles is subdiffusive~\cite{Shepelyansky:DisorderNonlin:PRL08,Flach:DisorderNonlin:PRL09,Flach:DisorderNonlineChaos:EPL10,Pikovsky:ScalingPropertiesWeakChaos:PRE11,Basko2011,Ivanchenko2011,Vermersch:AndersonInt:PRE12}.
These studies pointed out in particular the central role of chaotic
dynamics in the destruction of AL.

This approach has however two important drawbacks. The first is that
the timescale of subdiffusion is larger by several orders of magnitude
than the timescale of the single-particle dynamics, implying very
long computer calculations which make difficult a full study of the
interplay between disorder and interactions. The second is that the
nonlinearity leads to a strong dependence on the initial conditions
which also makes it difficult to give a {}``global'' characterization
of the different dynamical regimes. In previous works, we have shown
the existence of scaling laws \emph{with respect to the width of the
initial state}, allowing, to some extent, such a global characterization~\cite{Vermersch:AndersonInt:PRE12},
and demonstrated that these scaling laws are robust with respect to
decoherence effects ~\cite{Vermersch2012a} (which can also destroy
AL).

In the present work we tackle the problem of interacting ultracold
bosons in a 1D disordered lattice using a spectral analysis that does
not require such long computational times, as the information on the
chaotic behaviour is {}``inscribed'' even in early times in the
spectrum of the dynamics. This allows us to perform a more complete
and precise study of the problem over a large range of parameters.
A central quantity in our study, the spectral entropy,  proves very
useful to characterize the dynamic behaviours, confirmed by comparing
the information extracted from the spectral entropy to the Lyapunov
exponent, another well-known measure of chaotic behaviour. Moreover,
we show that such behaviours are described by {}``log-normal'' laws
which can be scaled with respect to the initial conditions, and we
propose a simple physical interpretation of our findings.

\section{The model}

We use here the discrete nonlinear Schr\"odinger equation with diagonal
disorder. It is essentially the same model used in previous works~\cite{Vermersch:AndersonInt:PRE12,Vermersch2012a},
so we only give an outline of it here.

The mean-field theory applied to ultracold bosons in an optical (ordered)
lattice leads to the so-called Gross-Pitaevskii equation: 
\[
i\hbar\dot{\phi}(x)=\left[\frac{p^{2}}{2m}+V(x)+g_{1D}N|\phi(x)|^{2}\right]\phi(x)
\]
where $\phi$ is the macroscopic wavefunction (or order parameter)
describing a Bose-Einstein condensate well below the critical temperature,
$V(x)=-V_{0}\cos(2k_{L}x)$ is the optical potential and $g_{\mathrm{1D}}$
the 1D coupling constant~\cite{Petrov:LowDimensionalTrappedGases:JP404}.
Tight-binding equations are obtained by decomposing the wavefunction
$\phi(x)=\sum_{n}c_{n}(t)w_{n}(x)$ onto the set of localized Wannier
functions of the first band $w_{n}(x)$ associated to the $n^{\mathrm{th}}$
lattice site:
\begin{equation}
i\dot{c}_{n}=v_{n}c_{n}-c_{n-1}-c_{n+1}+g\left|c_{n}\right|^{2}c_{n},\label{eq:DANSE}
\end{equation}
where we kept only (symmetric) nearest-neighbours couplings, which
is justified if the Wannier functions are strongly localized, that
is, for large enough $V_{0}$. According to usual conventions, we
measure distances in steps of the lattice and write energies in units
of the coupling constant of neighbour sites $T=-\int dxw_{n}(x)[p^{2}/2m+V(x)]w_{n+1}(x)$
\footnote{Wannier functions have the translation property $w_{n}(x)=w_{0}(x-n)$.%
}. Finally, times are written in units of $\hbar/T$. The coefficient
$v_{n}=\int dxw_{n}(x)[p^{2}/2m+V(x)]w_{n}(x)/T$ is the diagonal
on-site energy and the effect of interactions is taken into account,
in a mean-field approach, by adding the nonlinear term $g\left|c_{n}\right|^{2}$
where the dimensionless interacting strength is
\[
g=\frac{g_{\mathrm{1D}}N\int d_{x}|w_{0}|^{4}}{T}.
\]
In the absence of disorder, the on-site energy $v_{n}$ does not depend
on the site index $n$ and can be thus set to $0$. According to the
Anderson's postulate~\cite{Anderson:LocAnderson:PR58}, we introduce
diagonal disorder by picking random on-site energies $v_{n}$ uniformly
in an interval $[-W/2,W/2]$.

For $g=0$, \eref{eq:DANSE} describes the standard Anderson model,
and we shall call the corresponding eigenstates (eigenvalues) {}``Anderson''
eigenstates (eigenvalues). A few facts about it will be useful in
what follows. The eigenstates are exponentially localized in average
$\overline{c_{n}^{\nu}}\sim\exp\left(-|n|/l_{\nu}\right)$ (the overbar
indicates the averaging over realizations of the disorder) with a
localization length $l_{\nu}(W)\sim96(1-\epsilon_{\nu}^{2}/4)/W^{2}$~\cite{MuellerDelande:DisorderAndInterference:arXiv10}.
For $g\neq0$ the equation becomes nonlinear, and it is useful, if
not strictly rigorous, to interpret the nonlinear term as a {}``dynamical
correction'' to the on-site energy $v_{n}^{\mathrm{NL}}=g|c_{n}|^{2}$.
Previous works~\cite{Shepelyansky:DisorderNonlin:PRL08,Flach:DisorderNonlin:PRL09,Flach:DisorderNonlineChaos:EPL10,Ivanchenko2011,Pikovsky:ScalingPropertiesWeakChaos:PRE11,Vermersch:AndersonInt:PRE12}
put into evidence the existence of three main dynamical regimes: For
$g\ll W$, Anderson localization is expected to survive for very long
times, a regime that we shall call {}``quasi-localized''. For $g\sim W$,
the nonlinear correction $v_{n}^{\mathrm{NL}}$ induces chaotic dynamics
leading to subdiffusion and the destruction of AL. For $g\gg W$,
the very large nonlinear term $v_{n}^{\mathrm{NL}}$ decouples all
sites whose populations are not nearly equal (even in the absence
of the disorder), suppressing the diffusive behaviour and leading
to another type of localization, called self-trapping. This dynamics
does not rely on quantum interference and is therefore very robust
against external perturbations, including decoherence~\cite{Vermersch2012a}. 

Our aim is to characterize the global dynamics on a relatively short
timescale. We thus study the evolution according to \eref{eq:DANSE}
of bosons in a 1D box containing $L$ sites (typically, $L=101$)
and put an exponential absorber at each end of the box in order to
prevent wavepacket reflection. The norm of the wavepacket is thus
not anymore conserved as soon as it {}``touches'' the borders, and
we characterize the diffusive behaviour by calculating the survival
probability $p(t)=\sum_{n}|c_{n}|^{2}$; a value $p(t)<1$ indicates
that the packet has diffused outside the box. As in ref.~\cite{Vermersch:AndersonInt:PRE12},
we restrict the analysis to initial wavepackets of the form\begin{equation*}
c_{n}(t=0)=\cases{L_0^{-1/2}\exp\left(i\theta_{n}\right)&$|n|\le\left(L_{0}-1\right)/2$\\0&overwise}
%\cases{L_0^{-1/2}\exp\left(i\theta_{n}\right)&$}
\end{equation*}with $L_{0}\ll L$ %
\footnote{As discussed in more detail in ref.~\cite{Vermersch:AndersonInt:PRE12},
this form of wavepacket has the advantage of, on the average, projecting
onto all Anderson eigenstates, thus rendering the dynamics roughly
independent of the wavepacket energy.%
}.

\begin{figure}
\begin{centering}
\includegraphics[width=0.4\columnwidth]{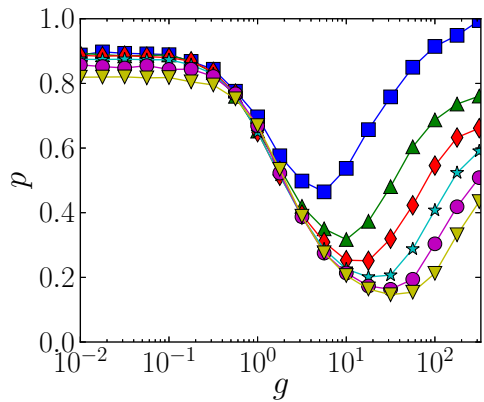}
\par\end{centering}

\caption{\label{fig:pvsg}Survival probability $p$ at time $t=10^{5}$ \emph{vs}
interaction strength $g$ for $W=4$ and different widths of the initial
state : $L_{0}=$3 (blue squares), 7 (green triangles), 13 (red diamonds),
21 (cyan stars), 31 (magenta circles), 41 (yellow inverted triangles).}
\end{figure}

The typical behaviour of the survival probability is illustrated in
\fref{fig:pvsg}, which represents $p(g,t=10^{5})$ as a function
of the interaction strength $g$ for $W=4$ and for various values
of $L_{0}$. To obtain such smooth curves, the survival probability
was averaged over typically $500$ realizations of the disorder and
of the initial phases $\theta_{n}$. Such long times are necessary
in order to clearly put into evidence the three different regimes
mentioned above, which are then easily identifiable: At low $g$,
the survival probability is close to one, corresponding to the quasi-localized
regime in which AL survives. For intermediate values of $g$, the
decrease of $p$ indicates that the wave packet has spread along the
box and part of it has been absorbed at the borders, corresponding
to a chaotic regime induced by the nonlinearity that destroys AL.
For large values of $g$, the survival probability increases again
and gets close to $1$, indicating that the packet is self-trapped~\cite{Shepelyansky:DisorderNonlin:PRL08,Flach:DisorderNonlin:PRL09,Vermersch:AndersonInt:PRE12}
and has never touched the borders if initially it was thin enough
(e.g. in the $L_{0}=3$ case). In the next section, we show that a
complementary way to describe the dynamics allows to a very precise
characterization within much shorter computation times.

\section{Characterizing chaotic dynamics with the spectral entropy\label{sec:SpecralEntropy}}

Spectral analysis is a very useful way of analyzing a chaotic dynamics,
be it in classical, {}``quantum'' %
\footnote{Traditionally the term {}``quantum chaos'' designates the behaviour
of a quantum (linear) system whose classical (nonlinear) counterpart
is chaotic.%
} or quantum nonlinear chaotic systems~\cite{Lepers:QuasiClassTrack:PRL08}.
The spectral entropy is a measure of the {}``richness'' of a spectrum.
Chaotic behaviours are associated to continuous spectra, and thus
to a high spectral entropy. Given a quantity $M(t)$, its power spectrum
is defined as
\[
S_{f}[M(t)]=\frac{\left|\tilde{M}(f)\right|^{2}}{\int_{0}^{f_{\max}}df\left|\tilde{M}(f)\right|^{2}}\quad,f\in[0,f_{\max}]
\]
where $\tilde{M}(f)$ is the Fourier transform of $M(t)$ for $t\in[0,t_{\max}]$.
The choice of the value of $t_{\max}$ determines the resolution of
the spectrum. In our case, we chose $t_{\max}=200$, which is a very
small value compared to the typical time of emergence of the interacting
regime ($t\sim10^{5}$) but which will be shown to be sufficiently
high to characterize the dynamics from a spectral point a view. As
we do not expect excitations whose timescale is inferior to the tunneling
time ($1$ in our rescaled units), we set $f_{\mathrm{max}}=1$ %
\footnote{None of our results is modified if we set $f_{\max}=2$.%
}. From the power spectrum, one defines the spectral entropy as

\begin{equation}
H=-\frac{\int df\ S(f)\log S(f)}{\log(f_{\mathrm{max}})}.\label{eq:H}
\end{equation}
For a perfectly monochromatic signal $S_{f}=\delta(f-f_{0})$, the
spectral entropy is zero, whereas for a white noise ($S_{f}=1/f_{\mathrm{max}}$)
$H=1$. The spectral entropy {}``counts'' the number of frequencies
which are present in the signal, and is a good indicator of the chaoticity
of the system~\cite{Rezek:StochasticComplexityMeasures:ITBE98}.
The spectral entropy obviously strongly depends on the choice of the
observable, and its usefulness as a dynamics indicator is reliant
on this choice.

A good observable in the present problem is the so-called {}``participation
number'' $P$ with respect to the Anderson eigenstates, which is
defined, in the present case, as follows. For a given realization
of disorder $\{v_{n}\}$, we calculate the Anderson eigenstates in
the Wannier basis, $\phi_{\nu}(x)=\sum_{n}d_{n}^{(\nu)}w_{n}(x)$
corresponding to an energy $\epsilon_{\nu}$, which are solutions
of \eref{eq:DANSE} with $g=0$ (the $d_{n}^{(\nu)}$ replacing the
$c_{n}$). Back to the $g\neq0$ case, \Eref{eq:DANSE} allows us
to calculate the evolution of the wavepacket $\psi(t)=\sum c_{n}(t)w_{n}(x)$
under the action of both disorder and nonlinearity%
\footnote{We use a standard Crank-Nicholson scheme with time-step $0.01<dt<0.1$.%
}. At any time, we can express the wavepacket in the Anderson eigenstates
basis
\begin{equation}
\psi(t)=\sum_{\nu}q_{\nu}(t)\phi_{\nu}(x),\label{eq:WavepacketAndersonEigeinbasis}
\end{equation}
from which, trivially, $q_{\nu}(t)=\sum_{n}d_{n}^{(\nu)}c_{n}(t)$.
The participation number is then defined as: 
\begin{equation}
P=\frac{\sum_{\nu}|q_{\nu}|^{2}}{\sum_{\nu}|q_{\nu}|^{4}}.\label{eq:ParticipationNb}
\end{equation}
 If $g=0$, the $\phi_{\nu}$ are the exact eigenstates of the problem,
so that the populations $|q_{\nu}|^{2}$ are constant. In the non-interacting
case $g\ne\text{0}$, the $|q_{\nu}|^{2}$ evolve under the action
of the nonlinearity. The participation number roughly {}``counts''
the number of Anderson eigenstates participating significantly in
the dynamics%
\footnote{To see this intuitively, consider two limit cases: If only one $\nu=\nu_{0}$
Anderson eigenstate is populated, $|q_{\nu}|^{2}=\delta_{\nu,\nu_{0}}$
and thus $P=1$; if $L_{0}$ eigenstates are equally populated, $|q_{\nu}|^{2}=L_{0}^{-1}$
and $P=L_{0}$. %
} and its time evolution thus reflects the apparition of Anderson eigenstates
that were not initially populated. 

\begin{figure*}
\centering{}\includegraphics[width=0.3\columnwidth]{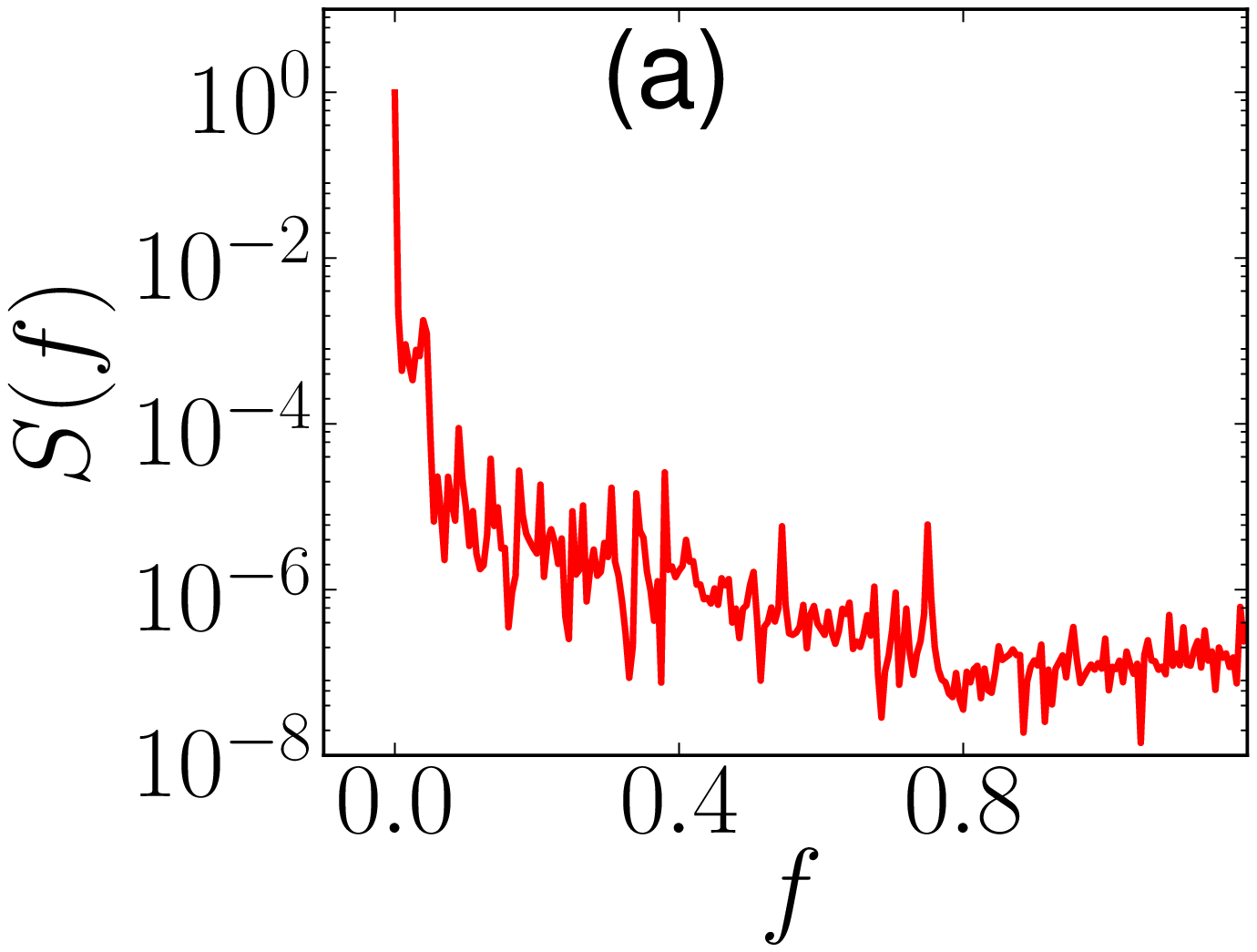}$\quad$\includegraphics[width=0.3\columnwidth]{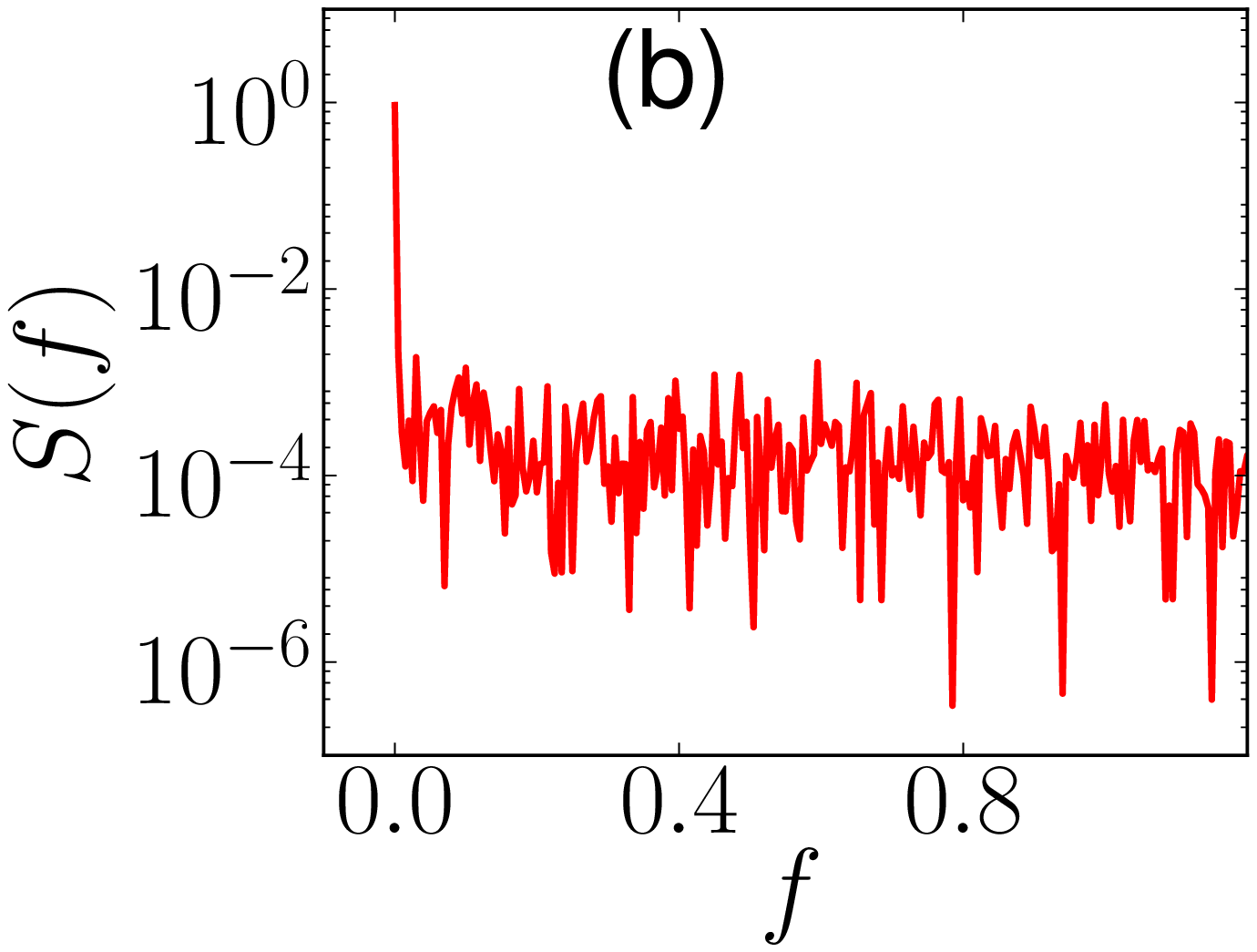}$\quad$\includegraphics[width=0.3\columnwidth]{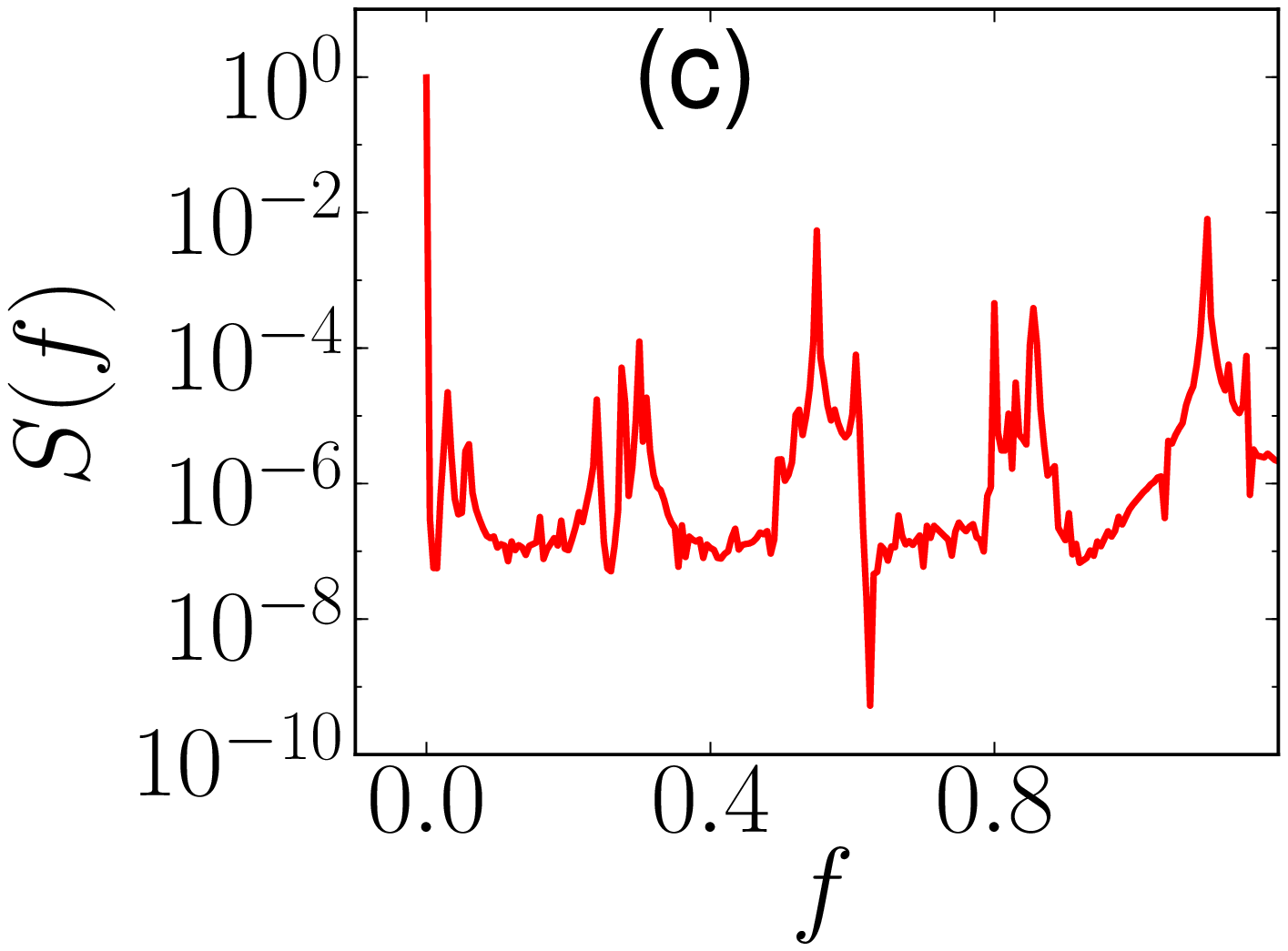}\caption{\label{fig:spectra}Example of spectral power $S(f)$ for the three
different regimes : a) localized $g=1$, b) chaotic $g=100$ and c)
self-trapped $g=1000$ . Other parameters are $W=3$, $L_{0}=3$ and
$t_{\max}=200$.}
\end{figure*}

\Fref{fig:spectra} shows the spectral power $S(f)$ of the participation
number $P$ in the three interacting regimes: (a) $g=1$ in the quasi-localized
regime, (b) $g=100$ in the chaotic regime, and (c) $g=1000$ in the
self-trapping regime. For $g=1$, the dynamics is very similar to
the linear case: The populations of Anderson eigenstates practically
do not evolve in time, as well as the participation number $P$ and
the spectrum is dominated by low frequencies. As a consequence, the
spectral entropy, calculated according to \eref{eq:H}, is relatively
small: $H=10^{-2}$. For $g=100$, Anderson eigenstates are strongly
coupled by the nonlinear term, each pair of coupled states generating
a Bohr frequency (shifted by the nonlinear correction), that is $\sim\epsilon_{\nu}+g\left|q_{\nu}\right|^{2}-\epsilon_{\nu'}-g\left|q_{\nu^{\prime}}\right|^{2}$.
In this regime, most Anderson eigenstates are coupled to each other,
so that the power spectrum is almost flat (with important local fluctuations)
at high frequencies, and the spectral entropy increases by almost
an order of magnitude $H=10^{-1}$ with respect to the preceding case.
For $g=1000$ the wavepacket is self-trapped and only a relatively
small number of Anderson eigenstates whose populations happen to be
close enough can interact. The spectrum is therefore dominated by
a finite number of frequencies and the spectral entropy is reduced,
$H=2\times10^{-2}$. However, although populations are stable in this
case, quantum phases may evolve chaotically under the action of the
nonlinearity~\cite{Thommen:ChaosBEC:PRL03}. Our definition of the
spectral entropy from the participation number -- which does not directly
depends on the phases -- excludes this {}``phase dynamics'' from
the corresponding spectrum.

\begin{figure}
\begin{centering}
\includegraphics[width=0.4\columnwidth]{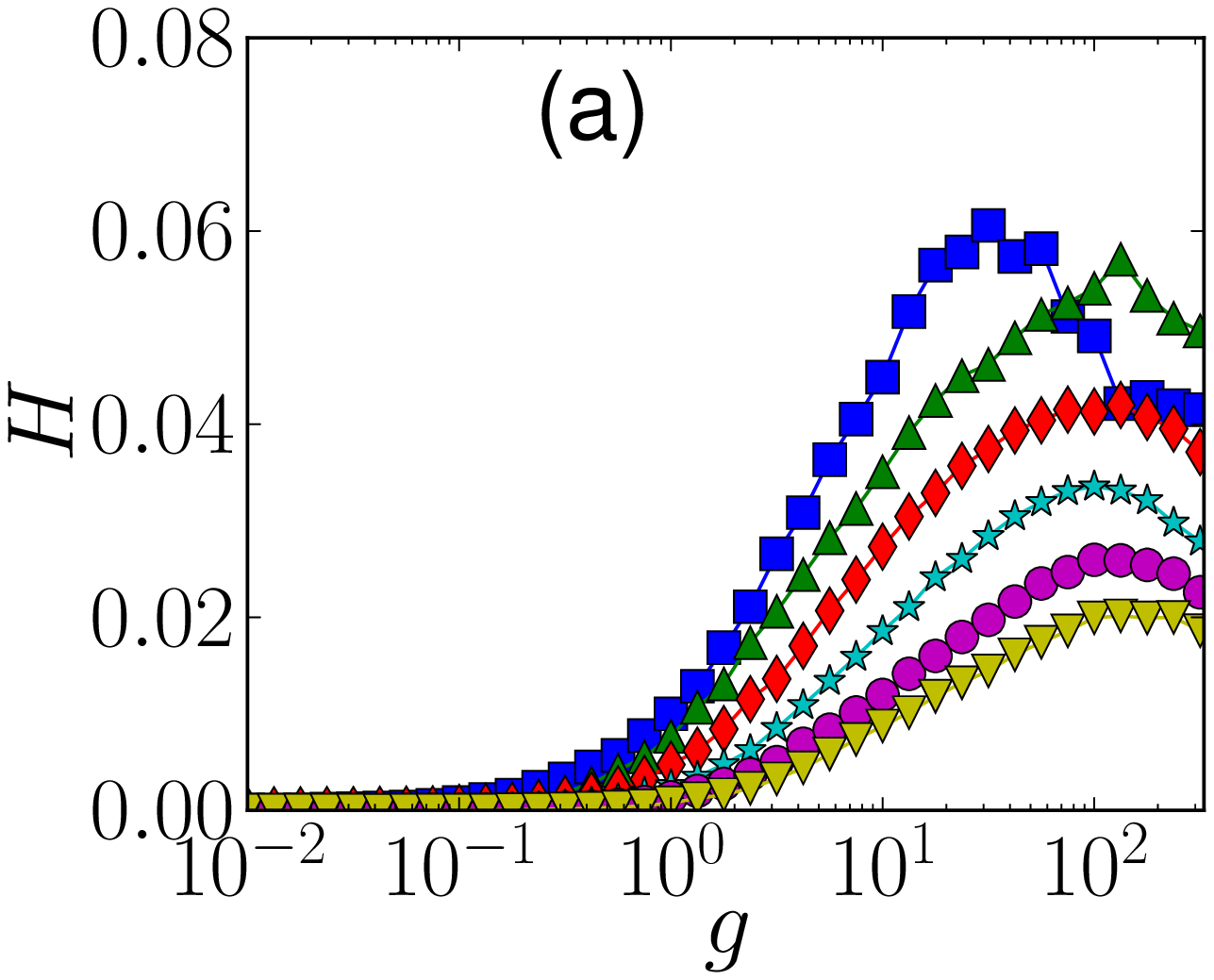}\qquad\includegraphics[width=0.4\columnwidth]{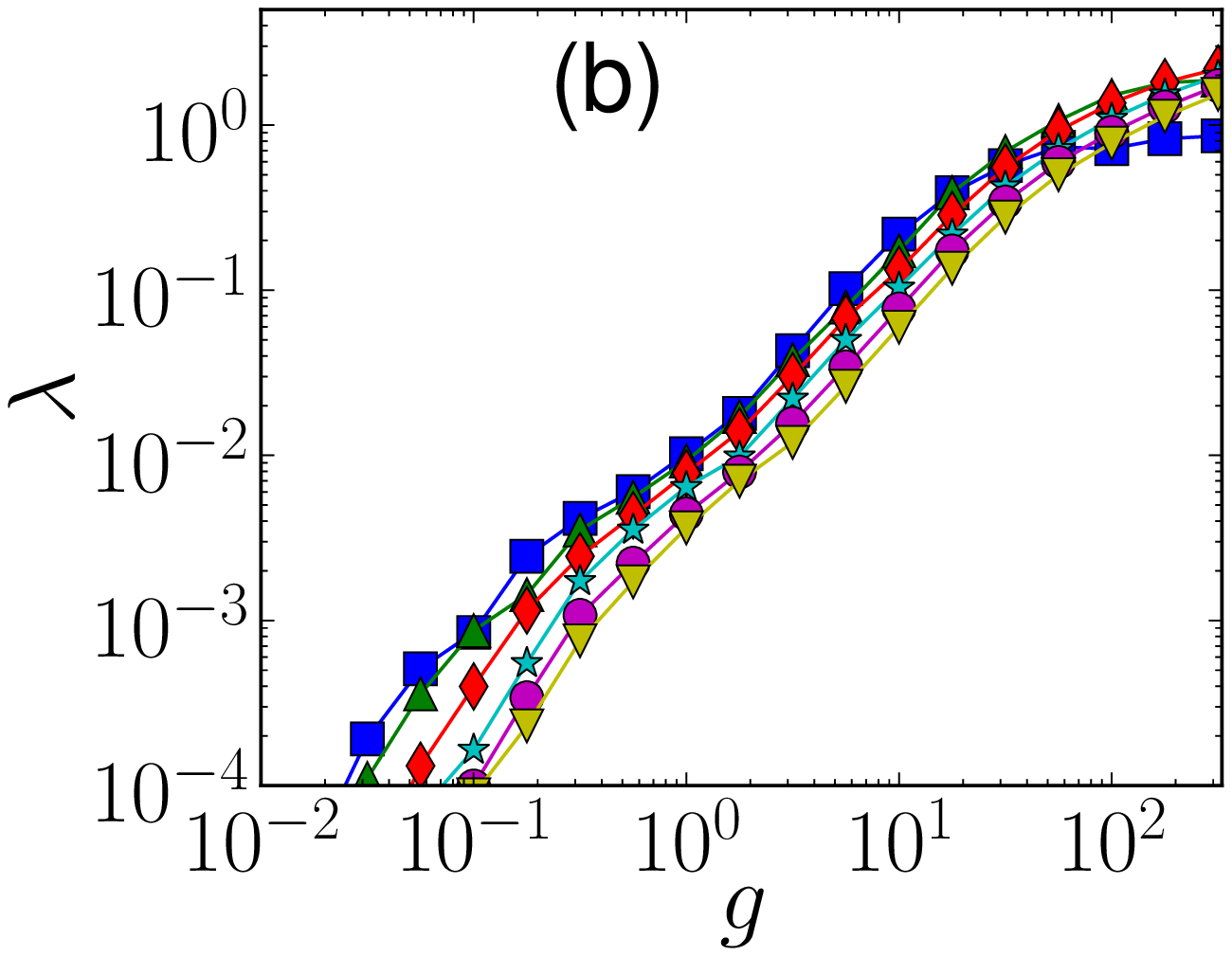}
\par\end{centering}

\caption{\label{fig:Avsg}(a) Spectral entropy $H$ and (b) Lyapunov exponent
$\lambda$ \emph{vs} interaction strength $g$ for $W=4$ and different
widths of the initial state : $L_{0}=$3 (blue squares), 7 (green
triangles), 13 (red diamonds), 21 (cyan stars), 31 (magenta circles),
41 (yellow inverted triangles).}
\end{figure}

We display in \fref{fig:Avsg}a the averaged spectral entropy as a
function of the nonlinearity parameter $g$ for different widths $L_{0}$
of the initial state. One clearly sees the crossover from the quasi-localized
to the chaotic regime, signalled by a marked increase of $H$. The
smaller the value of $L_{0}$, the smaller the value of the crossover.
This is easily understandable, as a more concentrated wavepacket leads
to a stronger nonlinear term $v^{\mathrm{NL}}$. On the right side
of the plot, one also sees, especially for low values of $L_{0}$
the beginning of a decrease of $H$ due to self-trapping.

It is interesting to compare the information obtained from the spectral
entropy to another relevant quantity characterizing chaos, the Lyapunov
exponent, which indicates how exponentially fast neighbour trajectories
diverge. This quantity is usually defined for classical systems, but
can be extended, with a little care, for quantum nonlinear systems~\cite{Lepers:QuasiClassTrack:PRL08}.
We present in \ref{app:Lyapunov} a method for calculating the Lyapunov
exponent of a quantum trajectory defined by amplitudes $c_{n}$ {[}\Eref{eq:DANSE}{]}.
\Fref{fig:Avsg}b displays the Lyapunov exponent $\lambda$ obtained
with the same parameters as in \fref{fig:Avsg}a. It displays a monotonous
increase with the nonlinear parameter, even in the region where the
spectral entropy decreases due to self-trapping, evidencing the presence
of a regime of {}``phase chaos'' mentioned above. This clearly shows
that $H$ and $\lambda$ provide different information on the dynamics
of the system, and that $H$ is clearly more adapted to distinguish
the three dynamical regimes.

In \sref{sec:LogNormalLawScaling} we will discuss the shape of these
curves, their scaling properties, and suggest a physical mechanism
explaining such properties.

\section{Log-normal law and scaling\label{sec:LogNormalLawScaling}}

As we shall see below, log-normal functions are ubiquitous in the
dynamics described in the present work, and so in various contexts
which are not obviously related to each other. A first example can
be found in \fref{fig:pvsg}, where each curve representing $p(\log g)$
can be fitted with a rather good accuracy by a inverted Gaussian function.
More generally, a log-normal function is defined as
\begin{eqnarray*}
f(x) & = & \frac{1}{x}\exp\left[-\frac{\left(\log x-\mu\right)^{2}}{2\sigma^{2}}\right]\\
 & = & \exp(-\mu+\frac{\sigma^{2}}{2})\exp\left[-\frac{(\log x-\mu+\sigma^{2})^{2}}{2\sigma^{2}}\right]
\end{eqnarray*}

In physics, such a function appears more often as a {}``log-normal
distribution'', related to the statistics of quantities which are
a \emph{product} of randomly distributed terms~\cite{Limpert:LogNormalDistibutions:BSC01}.
Log-normal statistics is very different from normal (Gaussian) statistics;
for example, the most probable value of a log-normally distributed
quantity is different from its average value. In the following, we
shall not only consider statistical distributions over the realizations
of the disorder: \fref{fig:pvsg} does not display the distribution
of $p$ over the realizations of disorder but its average as a function
of the interacting strength, so does \fref{fig:Avsg}a for the spectral
entropy. However and more interestingly, we now show that the statistical
distribution of the spectral entropy and of the Lyapunov exponent
are indeed log-normal.

\begin{figure}
\begin{centering}
\includegraphics[width=0.4\columnwidth]{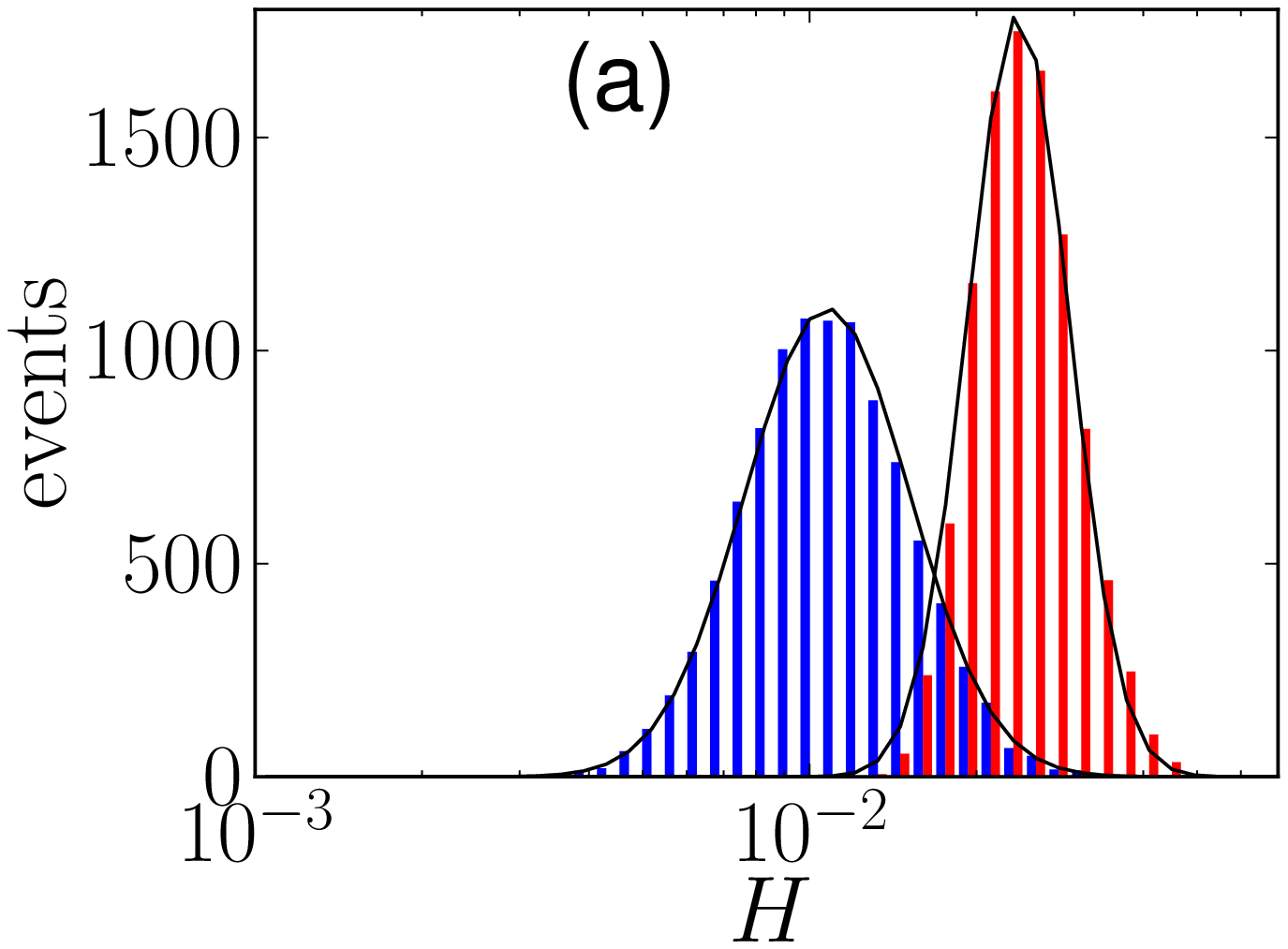}\qquad\includegraphics[width=0.4\columnwidth]{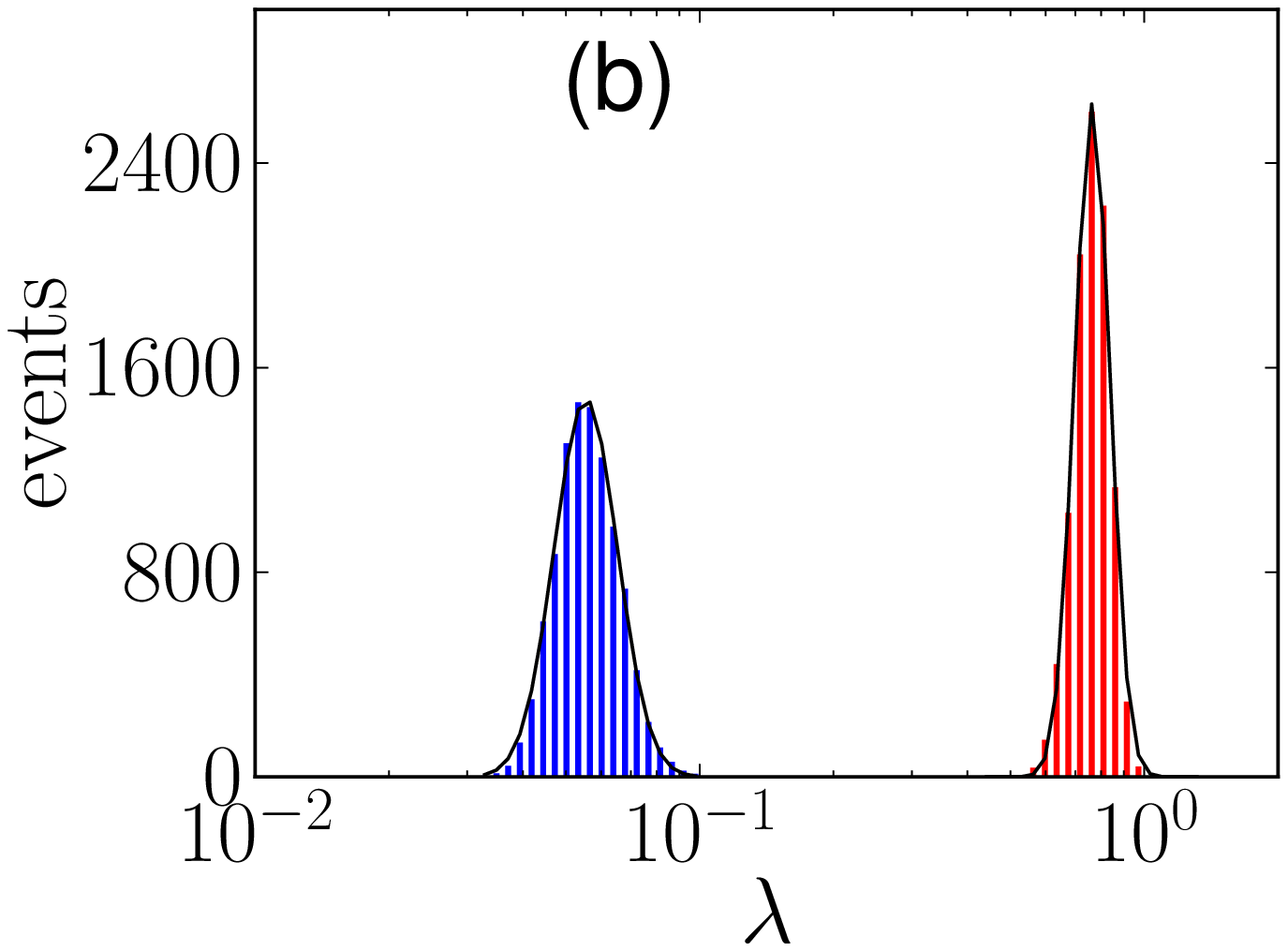}
\par\end{centering}

\caption{\label{fig:histograms}Histograms of the spectral entropy $H$ (a)
and the Lyapunov exponent $\lambda$ (b) for $W=3$, $L_{0}=41$ and
two interacting strengths : $g=10$ (left hand-side blue histogram),
$g=100$ (right hand-side red histogram). Black solid lines correspond
to log-normal fits.}
\end{figure}

Let us thus study the \emph{distribution} of these two quantities
over the realizations of the disorder $v_{n}$ and of the initial
phases $\theta_{n}$. \Fref{fig:histograms}a displays the distributions
of values of $H$ over $10^{4}$ realizations of the disorder for
two values of $g$, and \fref{fig:histograms}b the corresponding
distribution for the Lyapunov exponent $\lambda$. As shown by the
black fitting lines, both curves are perfectly fitted by a log-normal
function.

In order to have an idea of the origin of these shapes, one can consider
a simple, heuristic model. The destruction of AL is due to the nonlinear
coupling between Anderson eigenstates. Initially unpopulated eigenstates,
which, in the absence of nonlinearity, would never be populated, can
be thus excited thanks to a nonlinear transfer of population. The
goal of the simple model developed below is to characterize the statistical
distribution of such excitations. Projecting the wavepacket in the
Anderson eigenbasis {[}cf.~\Eref{eq:WavepacketAndersonEigeinbasis}{]},
\eref{eq:DANSE} then reads: 
\begin{equation}
i\dot{q}_{\nu}=\epsilon_{\nu}q_{\nu}+g\sum_{\nu_{1},\nu_{2},\nu_{3}}q_{\nu_{1}}^{*}q_{\nu_{2}}q_{\nu_{3}}I(\nu,\nu_{1},\nu_{2},\nu_{3})\label{eq:DANSE-nu}
\end{equation}
with $I(\nu,\nu_{1},\nu_{2},\nu_{3})=\sum_{n}d_{n}^{(\nu)}d_{n}^{(\nu_{1})}d_{n}^{(\nu_{2})}d_{n}^{(\nu_{3})}$.
Populations exchanges are controlled by the overlap $I$ of four coefficients.
The population transfer from eigenstate $\mu$ to eigenstate $\nu$
depends on $J_{1}=I(\mu,\mu,\nu,\nu)$, $J_{2}=I(\mu,\mu,\mu,\nu)$
and $J_{3}=I(\mu,\nu,\nu,\nu)$ , the term due to $J_{1}$ being for
instance : 

\[
\frac{d\left|q_{\nu}\right|^{2}}{dt}=2gJ_{1}\,\mathrm{Im}\left(q_{\nu}^{*2}q_{\mu}^{2}\right)
\]
In order to evaluate the probability distribution of $J_{1}$, we
make the assumption that the coupled Anderson eigenstates are exponentially
localized with the same localization length $\xi$. The exponential
localization is valid on the average and the localization length is
the same if the eigenstates have close enough eigenenergies. We thus
write 
\begin{eqnarray*}
d_{n}^{(\nu)}{}^{2} & = & \tanh\left(\frac{1}{\xi}\right)\exp\left(-\frac{2|n|}{\xi}\right)\\
d_{n}^{(\mu)}{}^{2} & = & \tanh\left(\frac{1}{\xi}\right)\exp\left(-\frac{2|n-l(\mu,\nu)|}{\xi}\right).
\end{eqnarray*}
The overlap sum $J_{1}$ can then be written as 
\begin{equation}
J_{1}=\tanh^{2}\left(\frac{1}{\xi}\right)e^{-2l(\mu,\nu)\text{/\ensuremath{\xi}}}\left[\frac{2}{1-e^{-4/\xi}}+l(\mu,\nu)-1\right]\label{eq:J}
\end{equation}
where, for simplicity, we have supposed that the spatial distance
between the eigenstates, $l(\mu,\nu)$, is an integer. The most important
term in \eref{eq:J} is $e^{-2l(\mu,\nu)/\xi}$. In the limit $\xi\to0$:
\[
J_{1}=e^{-2l(\mu,\nu)/\xi}\left[l(\mu,\nu)+1\right]
\]
The inverse localization length $\Lambda=1/\xi$ follows a \emph{normal}
distribution~\cite{Starykh:DynLocCavities:PRE00,MuellerDelande:DisorderAndInterference:arXiv10}
\[
P(\Lambda)\propto\exp[-(\Lambda-\Lambda_{0})^{2}/2\sigma^{2}]
\]
we obtain for the distribution of values of the overlap $J_{1}$
\begin{eqnarray*}
P(J_{1}) & = & P(\Lambda)\left|\frac{d\Lambda}{dJ_{1}}\right|\\
 & \propto & \frac{1}{J_{1}}\exp\left[-\frac{(\Lambda-\Lambda_{0})^{2}}{2\sigma^{2}}\right]\\
 & \propto & \frac{1}{J_{1}}\exp\left[-\frac{\left(\log J_{1}-G\right)^{2}}{2\tilde{\sigma}^{2}}\right]
\end{eqnarray*}
where $G=\log\left[l(\mu,\nu)+1\right]-2l(\mu,\nu)\Lambda_{0}$ and
$\tilde{\sigma}^{2}=4l(\mu,\nu)^{2}\sigma^{2}$. The overlap sum $J_{1}$
thus obeys a log-normal distribution.

We conjecture that overlap sums like $J_{1}$, controlling the coupling
between Anderson eigenstates, in fact control the destruction of the
Anderson localization, and thus explain the log-normal shapes we observed
for $H$ and $\lambda$. The above heuristic argument undoubtedly
presents various assumptions that are not rigorously justified, but
it has the merit of putting into evidence the intimate relation between
the log-normal distribution and the exponential localization of the
Anderson eigenstates. The link between the exponential shape and the
emergence of log-normal statistics has also been studied in the case
of the conductance of disordered systems~\cite{vanLangen:DisorderedWaveguide:PRE96,Evers:AndersonTransitions:RMP08,MuellerDelande:DisorderAndInterference:arXiv10}.

\begin{figure}
\centering{}\includegraphics[width=0.4\columnwidth]{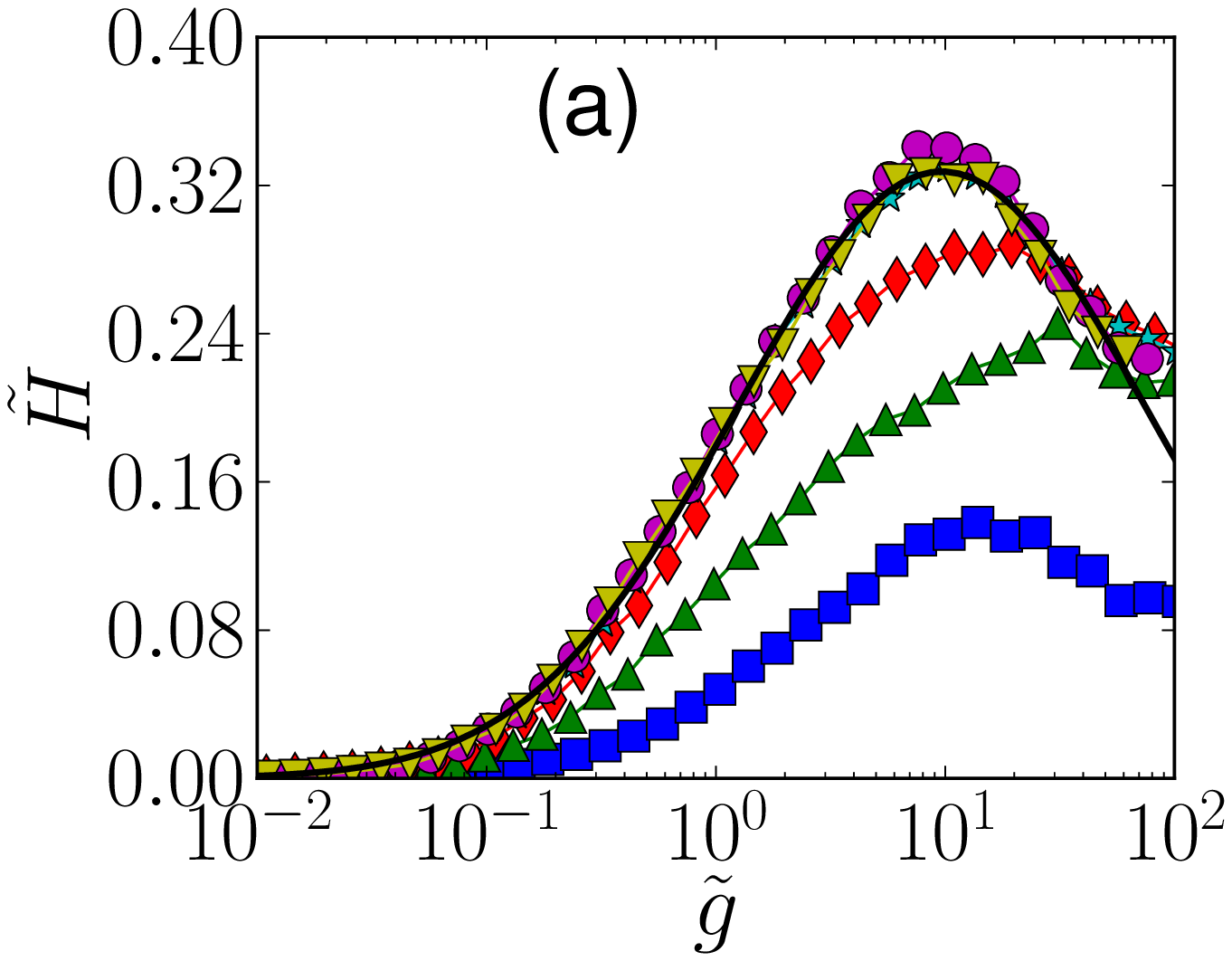}\qquad\includegraphics[width=0.4\columnwidth]{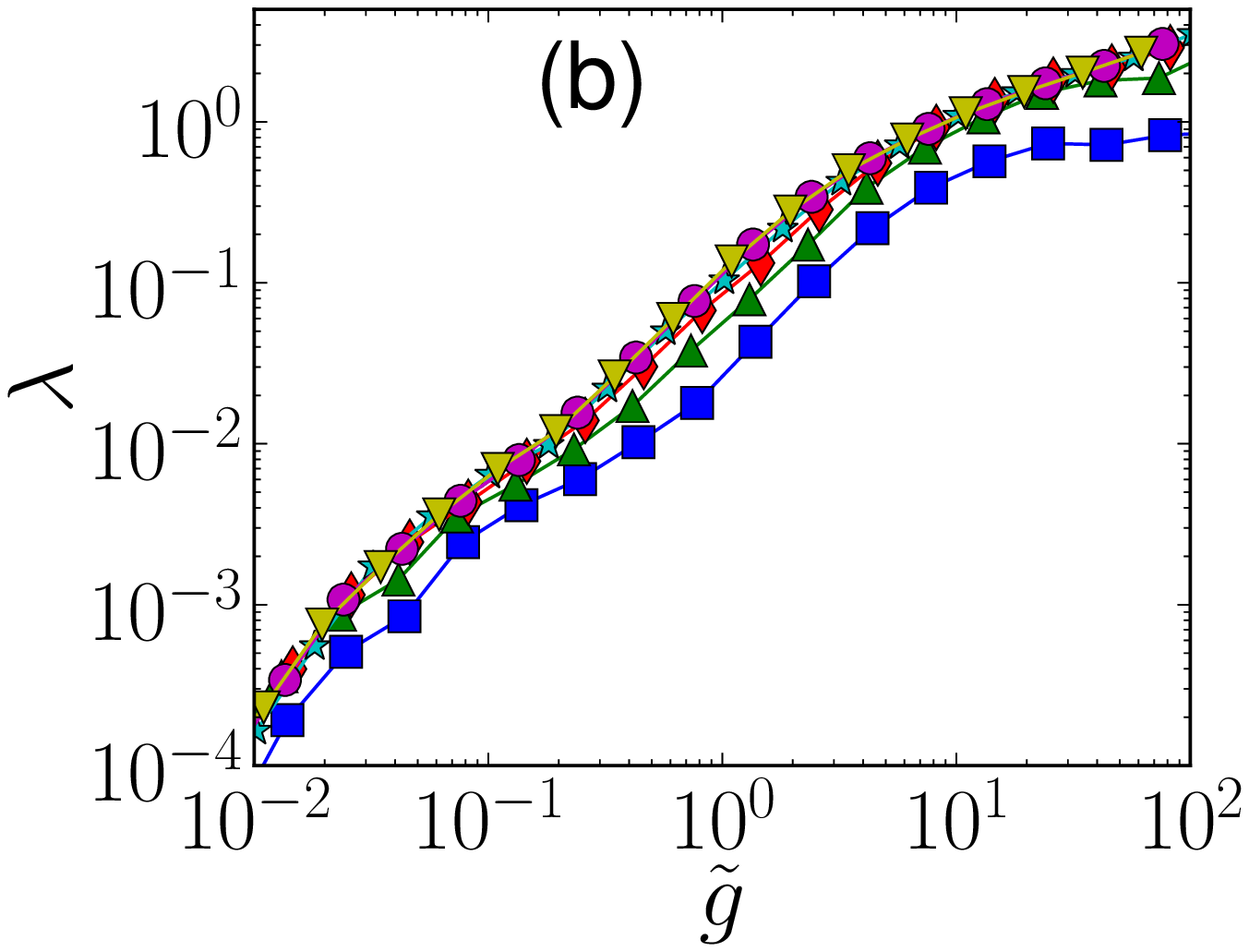}\caption{\label{fig:Atvsgt} (a) rescaled spectral entropy $\tilde{H}=HL_{0}^{s}$
with $s\approx3/4$ and (b) Lyapunov exponent $\lambda$ \emph{vs}
the rescaled interaction strength $\tilde{g}$ for $W=4$ and for
$L_{0}=$3 (blue squares), 7 (green triangles), 13 (red diamonds),
21 (cyan stars), 31 (magenta circles), 41 (yellow inverted triangles).}
\end{figure}

Let us now consider the averaged spectral entropy. In previous works~\cite{Vermersch:AndersonInt:PRE12,Vermersch2012a},
we showed that suitable scaling with respect to the initial state
width $L_{0}$ allowed a classification of dynamic regimes independently
of the shape of the initial state. In particular the interacting strength
$g$ was scaled as $\tilde{g}=gL_{0}^{-s}$ with $s\approx3/4$. This
scaling is meaningless for low values of $L_{0}$ because in this
case, the initial participation number is not of the order of $L_{0}$
but is of the order of the maximum Anderson localization length $\ell_{0}(W)\sim96/W^{2}$.
\Fref{fig:Atvsgt}a shows that scaling with $\tilde{g}$ and $\tilde{H}=HL_{0}^{s}$
make the curves $H(g)$ corresponding to $L_{0}\gtrsim20$ collapse
to a single curve, except in the strong self-trapped region. \Fref{fig:Atvsgt}b
shows the Lyapunov exponent as a function of $\tilde{g}$; the curves
also collapse for $L_{0}\gtrsim12$. The Lyapunov exponent itself
is independent of $L_{0}$ (with no scaling of the variable $\lambda$),
which is not surprising as it does not measure the absolute distance
between two quantum trajectories but the timescale of their exponential
divergence.

\begin{figure*}
\centering{}\includegraphics[height=0.15\textheight]{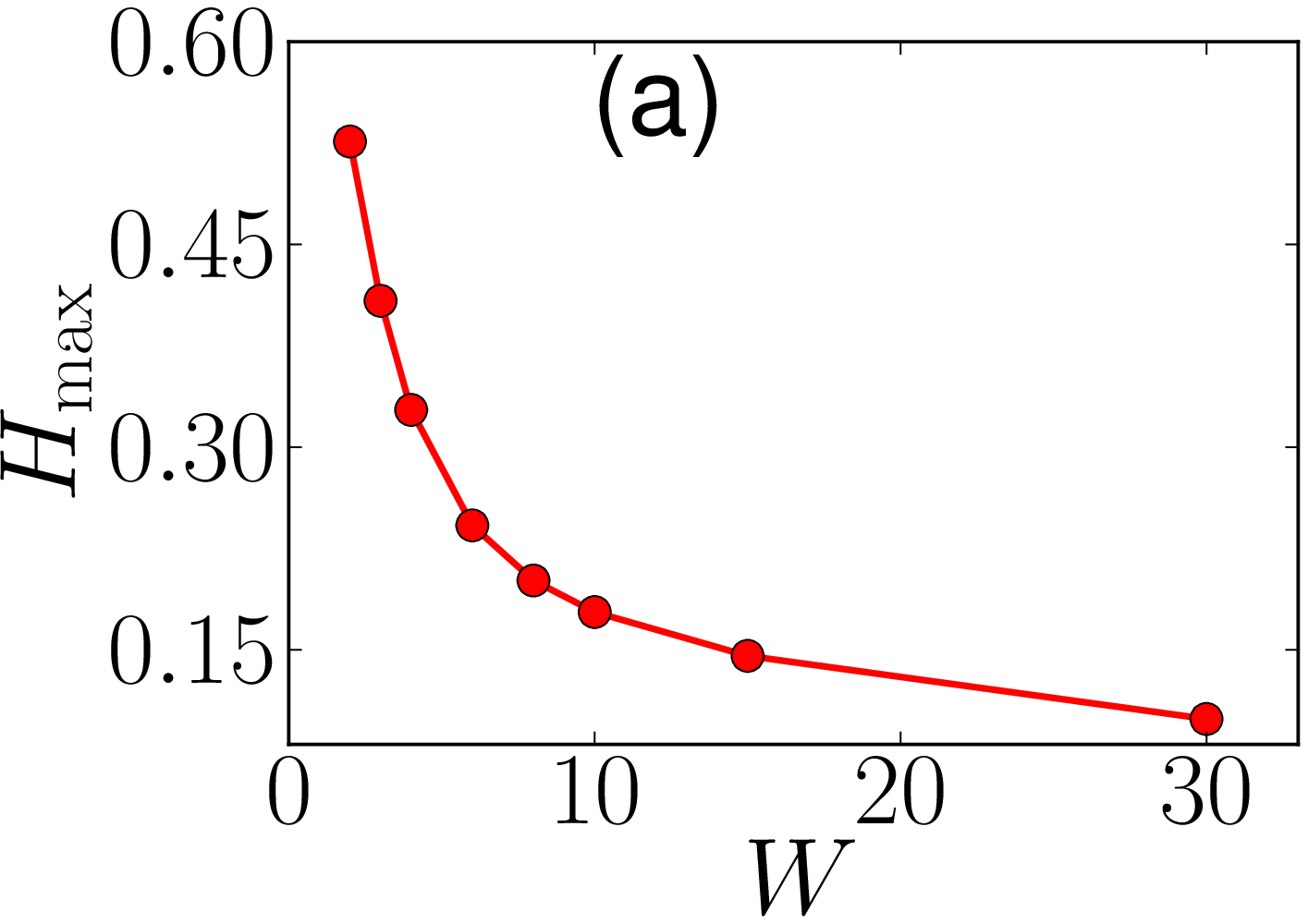}$\quad$\includegraphics[height=0.15\textheight]{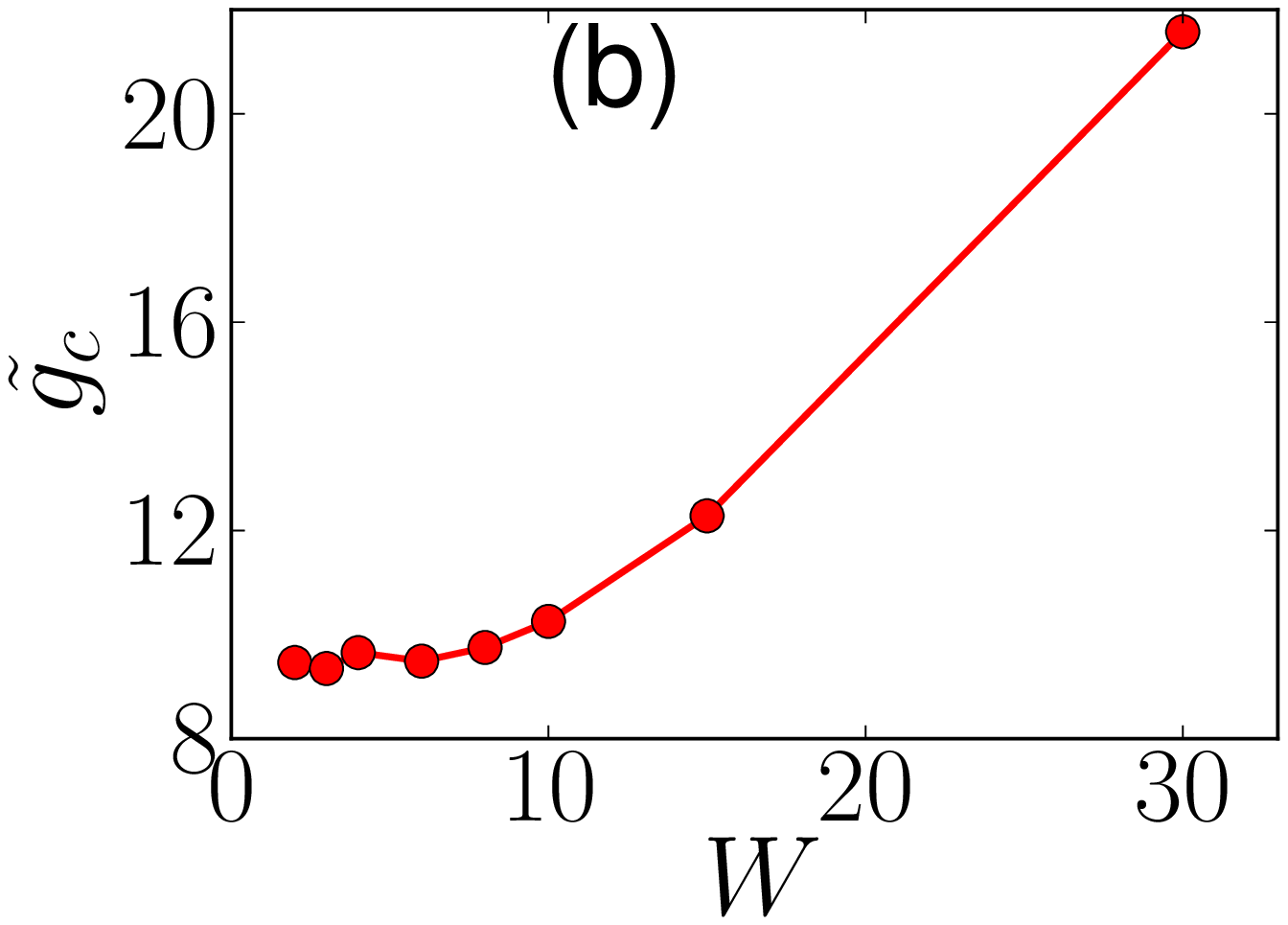}$\quad$\includegraphics[height=0.15\textheight]{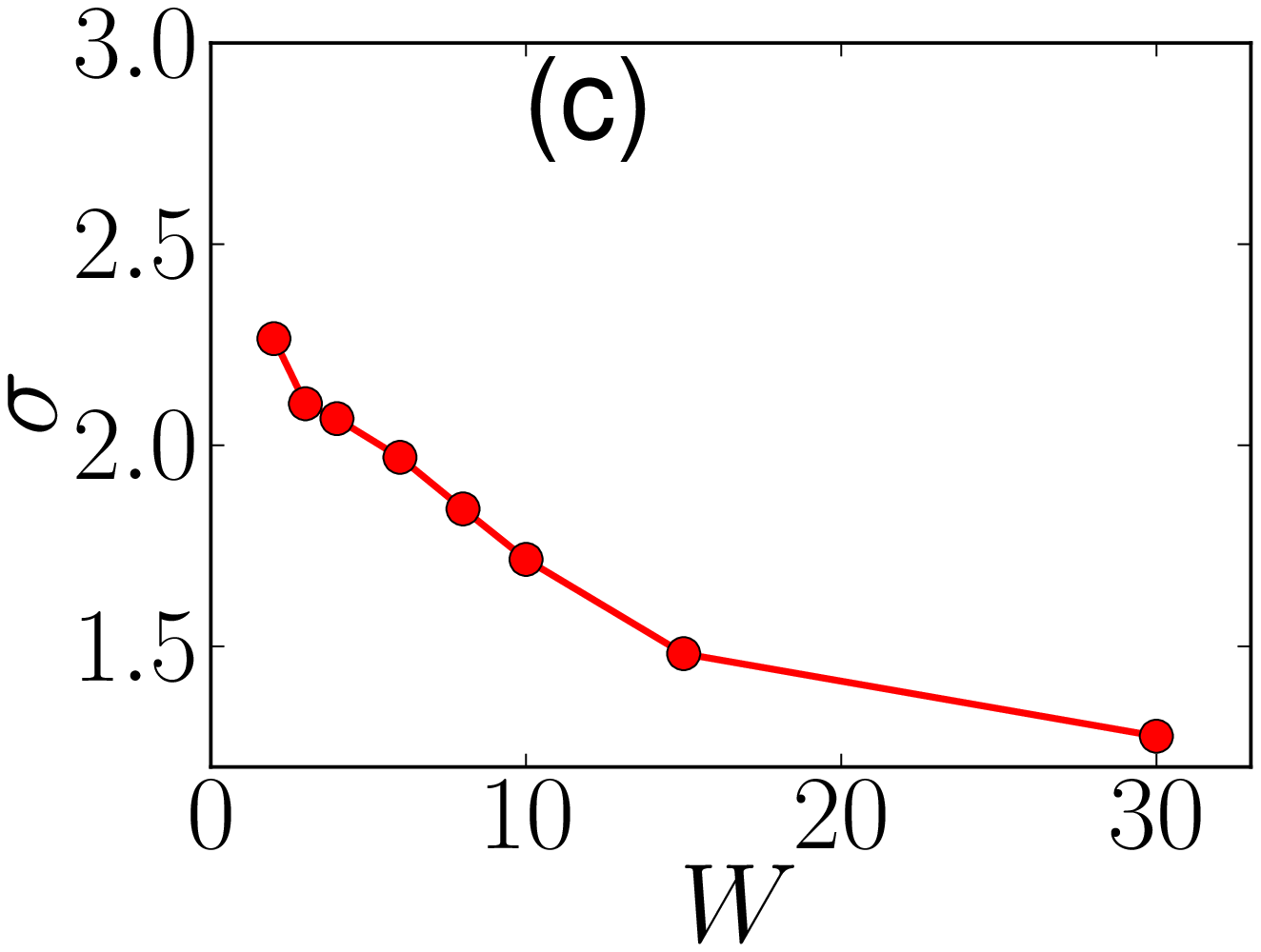}\caption{\label{fig:fit}a) Maximum entropy $\tilde{H}_{\max}$, b) corresponding
interaction strength $\tilde{g}_{\mathrm{c}}$ and c) resonant width
$\sigma$ vs disorder $W$.}
\end{figure*}

As shown by the black solid line in \fref{fig:Atvsgt}a, the scaled
spectral entropy $\tilde{H}$ is very well fitted (outside the strong
self-trapped region $\tilde{g}>100$) by the log-normal law:
\begin{equation}
\tilde{H}=\frac{h_{0}}{\tilde{g}\sqrt{2\pi\sigma^{2}}}\exp\left[-\frac{(\log\tilde{g}-G)^{2}}{2\sigma^{2}}\right]\label{eq:HtildeLogNorm}
\end{equation}
with three free parameters, the amplitude $h_{0}$, the center $G$
of the distribution and the {}``standard deviation'' $\sigma$.
The study of these fitting parameters as a function of the disorder
$W$ provides a full characterization of the dynamic regime. Instead
of representing the fit parameters $h_{0}$ and $G$, we prefer to
use more physical quantities, namely the maximum value of $\tilde{H}$,
$\tilde{H}_{\max}=\left[h_{0}\left(2\pi\sigma^{2}\right)^{-1/2}\right]\exp\left(\sigma^{2}/2-G\right)$
(\fref{fig:fit}a) and the rescaled interaction strength $\tilde{g}_{c}=\exp\left(G-\sigma^{2}\right)$
(\fref{fig:fit}b) corresponding to this maximum. The dependence of
the standard deviation $\sigma$ on $W$ is displayed in \fref{fig:fit}c.
The quantity $\tilde{H}_{\max}$ is a decreasing function of $W$:
as the localization length decreases with $W$, so does the overlap
between two neighbour Anderson eigenstates. On the contrary, $\tilde{g}_{\mathrm{c}}$
is an increasing function of $W$: The number of resonances is maximum
when $v_{n}^{\mathrm{NL}}$ is comparable to the typical energy between
two neighbour states, itself of the order of the bandwidth $\sim4+W$;
$\tilde{g}_{\mathrm{c}}$ is thus independent of $W$ at low disorders
and increases with $W$ for larger disorders. Finally, in \fref{fig:fit}c,
one can notice that the log-normal curve becomes sharper when the
disorder increases, which can be attributed to the fact that Anderson
Localization is more robust against interactions at high disorders.

\begin{figure}
\begin{centering}
\includegraphics[width=0.4\columnwidth]{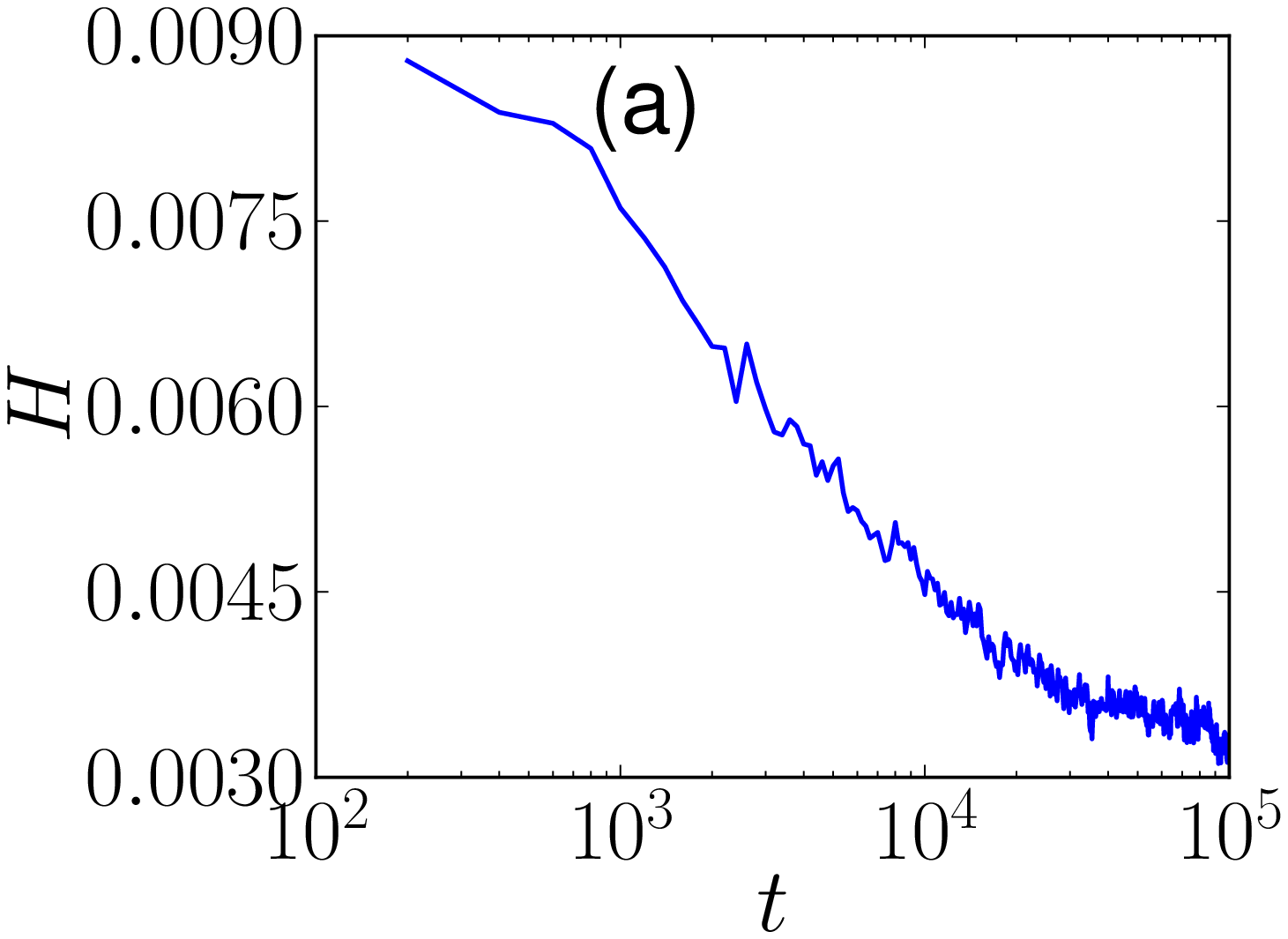}\qquad\includegraphics[width=0.4\columnwidth]{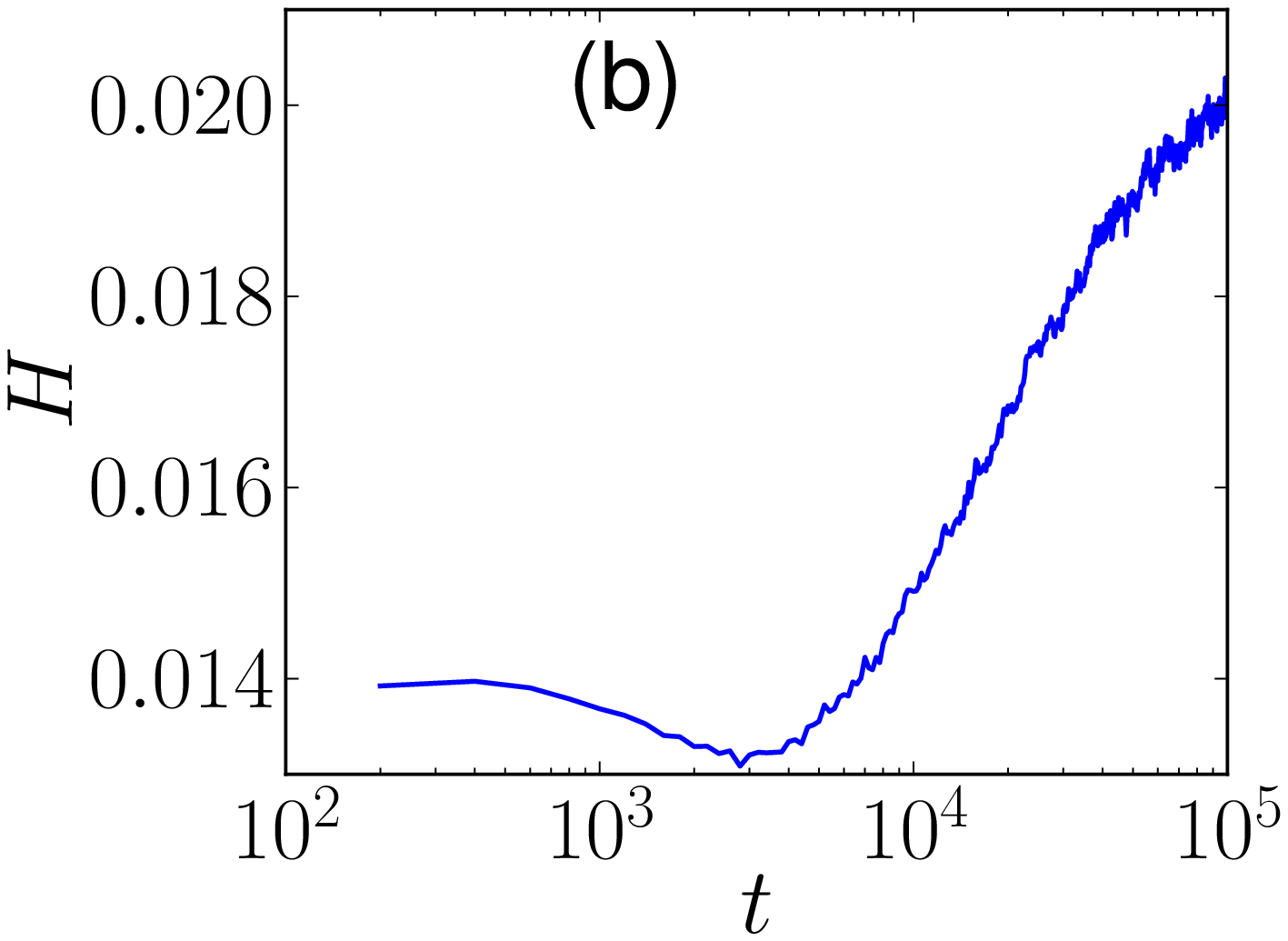}
\par\end{centering}

\caption{\label{fig:Hvst}Evolution of the spectral entropy $H(t)$ for $W=4,$
$L_{0}=41$. For (a) $g=10$ one essentially sees the effect of dilution,
which progressively weakens the chaoticity of the system. For (b)
$g=10^{3}$, there is a first phase, in which the self-trapping is
progressively destroyed which is followed by a slow transition towards
the chaotic regime.}

\end{figure}

The fact that the spectral entropy can be determined from a relatively
short time interval also allows one to follow the \emph{evolution}
of the dynamics. \Fref{fig:Hvst} shows the evolution of the spectral
entropy $H(t)$, defined as the spectral entropy calculated in an
interval $[t,t+200]$ , that we shall call, for short, the \emph{dynamic
spectral entropy}. For the low value of $g=10$ (\fref{fig:Hvst}a),
after the destruction of AL, the diffusion of the wavepacket produces
a dilution and a consequent diminution of the nonlinearity that reduces
the chaoticity, and thus the spectral entropy of the system. For the
high value of $g=1000$ (\fref{fig:Hvst}b) one sees a more complex
interplay of different regimes: The initial state is initially frozen
in a self-trapping regime, but this regime is unstable (for this particular
set of parameters): If a fraction of the packet escapes the self-trapped
region, the nonlinear contribution $v_{n}^{\mathrm{NL}}$ decreases
which leads to a weakening of the trapping. The destruction of the
self-trapping regime takes a much longer time that the destruction
of the AL. One first observes a small decrease of the spectral entropy,
as some eigenstates leave the box and do not interact anymore. Then,
when $v_{n}^{\mathrm{NL}}$ has decreased sufficiently in the center
of the packet, the system enters the chaotic regime.

\begin{figure*}
\begin{centering}
\includegraphics[height=0.15\textheight]{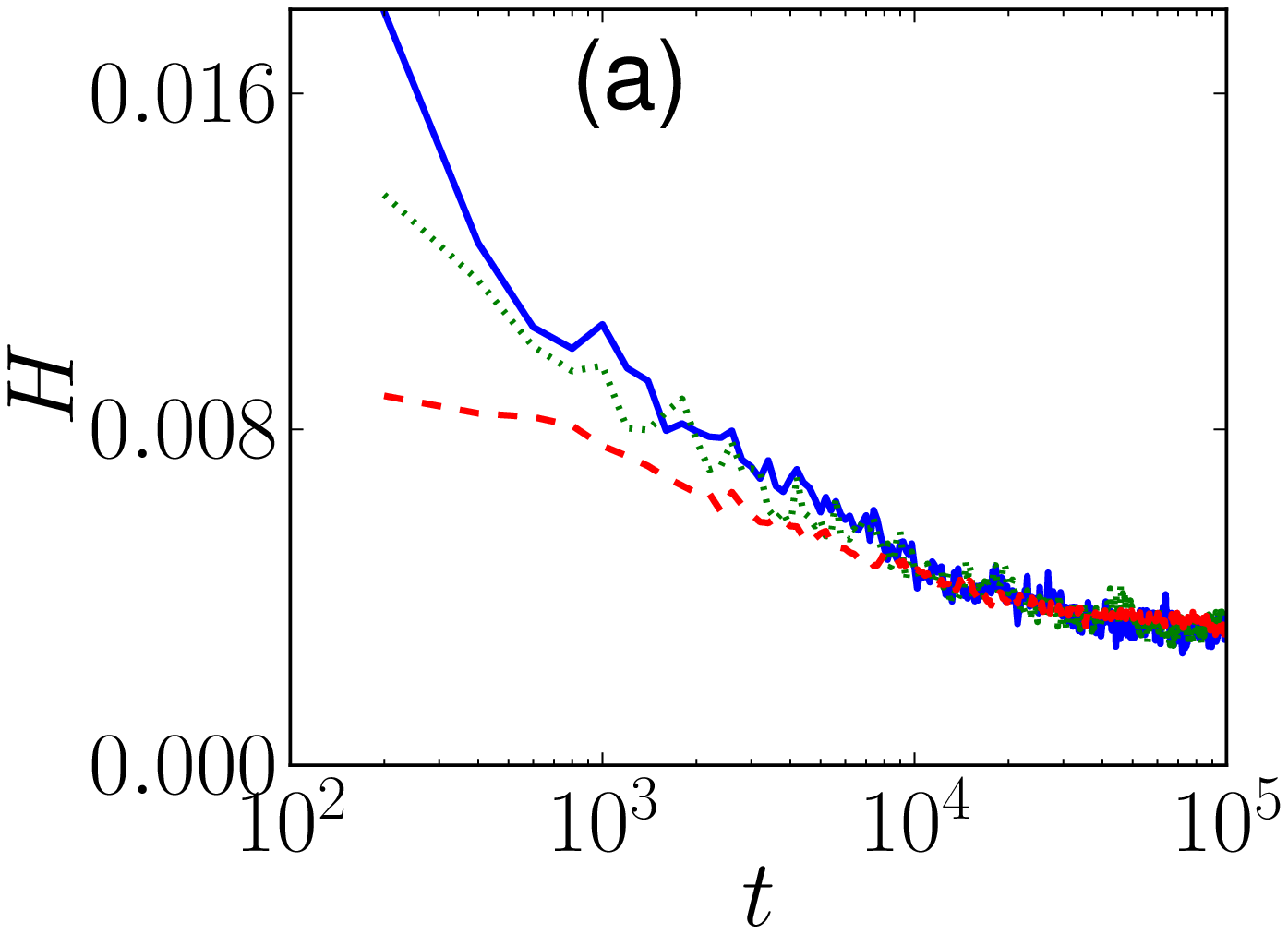}$\quad$\includegraphics[height=0.15\textheight]{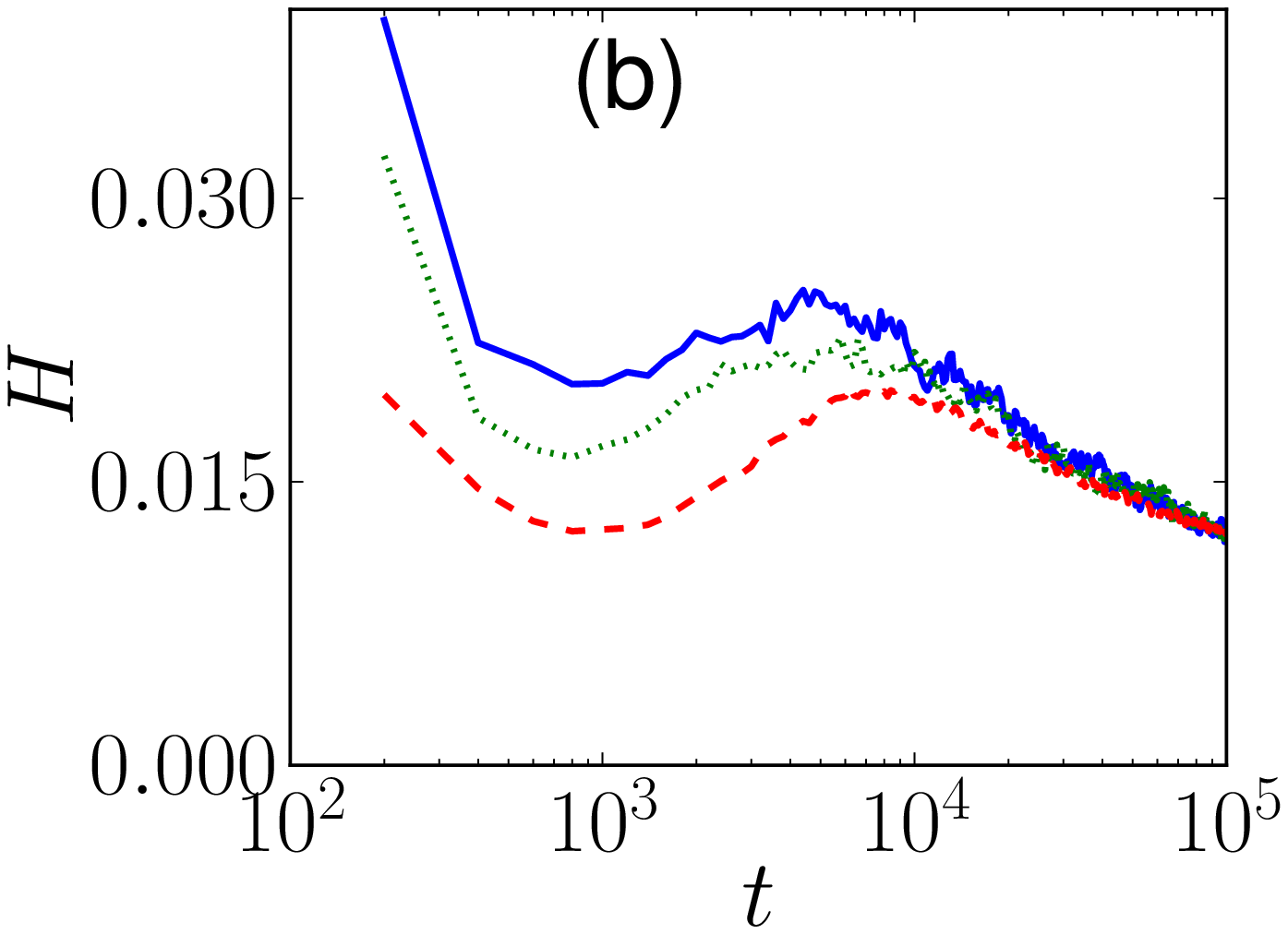}$\quad$\includegraphics[height=0.15\textheight]{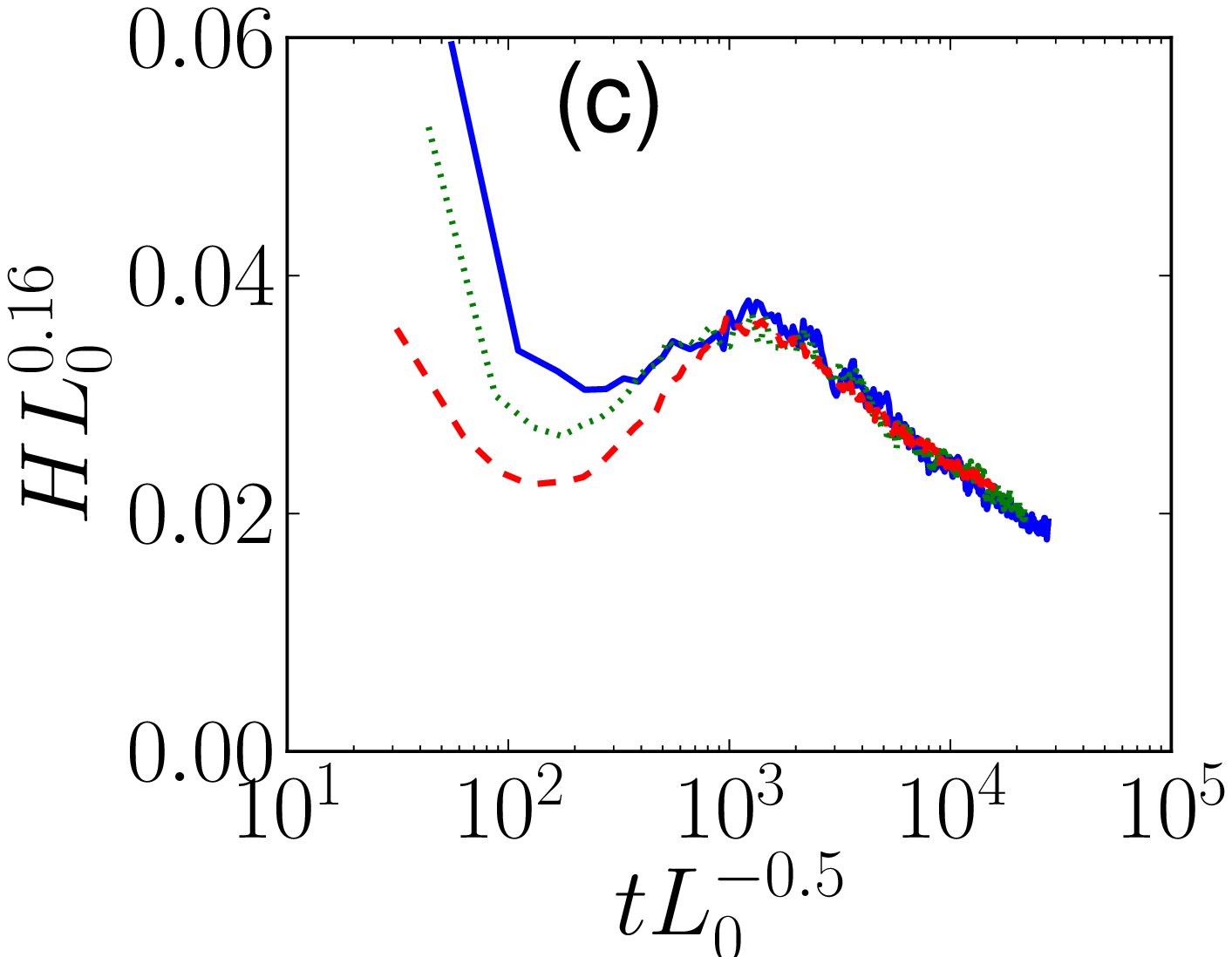}
\par\end{centering}

\caption{\label{fig:Htscaling} Scaling of the spectral entropy $H(t)$ for
$W=4$ and $L_{0}=13$ (blue full line), $L_{0}=21$ (green dotted)
and $L_{0}=41$ (red dashed). The use of the same values of the scaled
nonlinearity $\tilde{g}=gL_{0}^{-s}$ {[}$\tilde{g}=0.62$ (a) and
$\tilde{g}=6.17$ (b){]} shows that there is a \emph{universal} asymptotic
regime. Comparison of plots (b) and (c) shows that additional \emph{adhoc}
scaling gives an even more universal character to this behaviour.}
\end{figure*}

In plots a and b of \fref{fig:Htscaling} we represented the dynamic
spectral entropy $H(t)$ for three values of $L_{0}$, corresponding
to a \emph{same} value of $\tilde{g}$ (0.62 in a, 6.17 in b and c).
The curves converge for long times, putting into evidence the existence
of a \emph{universal} asymptotic regime, independent of the initial
state, once the proper scaling on $\tilde{g}$ is applied. Additional
scaling can even be used to describe the transition from the self-trapping
regime to the chaotic regime, as shown in \fref{fig:Htscaling}c.
Although we have presently no physical explanation for the exponents
appearing in these scaling laws, these results strongly support the
idea that the features of the nonlinear dynamics should be scaled
again with respect to the initial state~\cite{Vermersch:AndersonInt:PRE12}.

\section{Conclusion}

We have shown that the spectral entropy is a very good indicator of
the chaoticity of the dynamics, and allows a very good characterization
of the dynamics regimes. Moreover, it can be calculated dynamically
and also gives information on the \emph{evolution} of the dynamics.
The Lyapunov exponent gives, in the present context, a less complete
characterization of the dynamics. It describes very well the progressive
destruction of the Anderson Localization by the onset of chaotic behaviour,
exactly as the spectral entropy, but it does not give information
in the self trapping regime, where {}``phase chaos'' is still present.

The argument presented in \sref{sec:LogNormalLawScaling} suggests
that the log-normal shape is linked to the exponential localization
of the Anderson eigenstates. It is a bit surprising to find this link
between the behaviour of a strongly nonlinear system and the eigenstates
of the corresponding linear system. From a mathematical viewpoint,
spectral analysis (in the sense of the determination of eigenvalues
and eigenvectors) cannot be directly applied to nonlinear systems,
and, presently, there is no alternative analytical method for analyzing
nonlinear systems. The results discussed in the present work and in
refs.~\cite{Vermersch:AndersonInt:PRE12,Vermersch2012a} indicate
that \emph{scaling over the initial state} is a promising tool for
the study of the emerging field of nonlinear quantum mechanics. In
this context, it is worth noting that we have presently no convincing
explanation for the precise values of the exponents appearing in these
scaling laws, and that we have considered only a particularly simple
family of initial states. Moreover, there is no experimental verification
of these scaling laws. It thus appears that a large field of investigations
is opened for both theoreticians and experimentalists in the near
future.

\ack{}{Laboratoire de Physique des Lasers, Atomes et Mol\'ecules is Unit\'e
Mixte de Recherche 8523 of CNRS. Work partially supported by Agence
Nationale de la Recherche's LAKRIDI grant and {}``Labex'' CEMPI.}

\appendix

\section{Lyapunov exponent of a quantum trajectory\label{app:Lyapunov}}

We are interested in calculating the Lyapunov exponent of the quantum
trajectory $c\equiv(c_{n})$. Considering two initial conditions $c^{a}$,
$c^{b}$, which are separated by an infinitesimal distance $d_{0}$,
the Lyapunov exponent measures the rate of their exponential divergence.
\begin{equation}
\lambda=\lim_{t\to\infty}\frac{1}{t}\log\frac{d(t)}{d_{0}}\label{eq:lambda}
\end{equation}
where $d(t)=|c^{b}(t)-c^{a}(t)|$. In principle, one could use directly
\eref{eq:lambda} but it turns out that the trajectories do not evolve
in an ensemble of infinite volume. As a consequence the distance $d(t)$
rapidly saturates and $\lambda$ tends to $0$. The common method
to get rid of this drawback is to let trajectories evolve during a
short time period $d_{t}$ and then evaluate the corresponding Lyapunov
exponent
\begin{equation}
\lambda_{0}=\frac{1}{d_{t}}\log\frac{d_{1}}{d_{0}}\label{eq:lambda2}
\end{equation}

where $d_{1}=|c^{b}(d_{t})-c^{a}(d_{t})|$. Before letting the system
evolve for another time interval $d_{t}$, we rescale trajectories
so that the distance between $a$ and $b$ is set to $d_{0}$ keeping
their relative orientation unchanged. Usually, people rescale the
second trajectory $\tilde{c}^{b}=c^{b}+\frac{d_{0}}{d_{1}}(c^{b}-c^{a})$%
\footnote{see for example\newline http://sprott.physics.wisc.edu/chaos/lyapexp.htm.%
} satisfying immediately both requirements. We then obtain $\lambda_{1},\lambda_{2},..$
by iterating the three-steps operation : (i) evolution during a short
time $d_{t}$ (ii) calculation of the corresponding Lyapunov exponent
from \eref{eq:lambda2} (iii) renormalization of the trajectories.
We finally deduce the final Lyapunov exponent : 
\[
\lambda=\lim_{N\to\infty}\frac{1}{N}\sum_{i=1}^{N}\lambda_{i}
\]
where $N$ represents the number of iterations. Unfortunately, the
crucial rescaling operation changes the norm of the second trajectory
$c_{b}$ : in the case of quantum systems, the procedure is therefore
based on non-physical states. Here we propose to modify the step (iii)
rescaling both trajectories using the following scheme : 
\begin{eqnarray*}
\tilde{c}^{a} & = & \alpha c^{a}+\beta c^{b}\\
\tilde{c}^{b} & = & \gamma c^{a}+\delta c^{b}
\end{eqnarray*}

where $(\alpha,\beta,\gamma,\delta)$ satisfy the following conditions%
\footnote{We calculate the Lyapunov exponent without using an absorber potential
so that the norm $|c|^{2}$ is a conserved quantity.%
} : 

\begin{eqnarray}
\tilde{c}^{b}-\tilde{c}^{a} & = & \frac{d_{0}}{d_{1}}\left(c^{b}-c^{a}\right)\label{eq:rescale1}\\
|\tilde{c}^{a}|^{2} & = & 1\label{eq:rescale2}\\
|\tilde{c}^{b}|^{2} & = & 1\label{eq:rescale3}
\end{eqnarray}

Given that $d_{1}=|c^{b}-c^{a}|$, the first equation is nothing but
the usual scaling condition which is used for classical systems. Projecting
\eref{eq:rescale1} on $c^{a}$ and on $c^{b}$, one immediately obtains
\begin{eqnarray*}
\gamma & = & \alpha-\frac{d_{0}}{d_{1}}\\
\delta & = & \beta+\frac{d_{0}}{d_{1}}
\end{eqnarray*}

Subtracting \eref{eq:rescale2} from \eref{eq:rescale3}, and using
$2\Re\langle c^{a}|c^{b}\rangle=2-d_{1}^{2}$, we obtain 
\[
\beta=\alpha-\frac{d_{0}}{d_{1}}
\]
and finally from \eref{eq:rescale2}, we can choose 
\[
\alpha=\frac{d_{0}}{2d_{1}}+\frac{1}{2}\sqrt{\frac{4-d_{0}^{2}}{4-d_{1}^{2}}}
\]
to finally deduce $\beta,\gamma,\delta$. This method allows an accurate
calculation of the Lyapunov exponent in the discrete system considered
in the present work, but it can in principle be also applied to the
continuous Gross-Pitaevskii equation.

\section*{References}

\bibliographystyle{unsrt}
\bibliography{ArtDataBase,library}

\end{document}